\documentclass[12pt,preprint]{aastex}

\newcommand{\myemail}{lic@nju.edu.cn}

\shorttitle{Coronal magnetic topology and SIEEs}
\shortauthors{Li et al.}

\begin{document}


\title{Coronal magnetic topology and the production of solar impulsive energetic electrons}


\author{C. Li\altaffilmark{1,2}, L. P. Sun\altaffilmark{1}, X. Y. Wang\altaffilmark{2,3}, and Y. Dai\altaffilmark{1,2}}
\altaffiltext{1}{School of Astronomy and Space Science, Nanjing University, Nanjing 210093, China. \myemail}
\altaffiltext{2}{Key Laboratory for Modern Astronomy and Astrophysics (Nanjing University), Ministry of Education, Nanjing 210093, China}
\altaffiltext{3}{Key Laboratory of Solar Activity, The National Astronomical Observatories, CAS, 20A Datun Road, Chaoyang District, 100012 Beijing, PR China}



\begin{abstract}
We investigate two candidate solar sources or active regions (ARs) in association with a solar impulsive energetic electron (SIEE) event on 2002 October 20. The solar particle release (SPR) times of SIEEs are derived by using their velocity dispersion with consideration of the instrumental effect. It is found that there are double electron injections at the Sun. The low-energy ($\lesssim$13 keV) electron injection coincides with a C6.6 flare in AR10154 and is accompanied with prominent type III radio bursts rather than a stronger M1.8 flare in AR10160. The M1.8 flare produces, however, faint type III radio bursts. Electrons of $\sim$25 to $\sim$300 keV are released $\sim$9 min later when a jet-like CME travels to $\sim$2.6 solar radii. We further examine the coronal magnetic configurations above the two ARs based on the potential field source surface (PFSS) model. It is found that open field lines, rooted in AR10154 and well connected to the Earth, provide escaping channels for energetic electrons. Only a small portion of magnetic fields are opened above AR10160, being responsible for the faint type III radio bursts. These lines are, however, not well connected, making it impossible for SIEEs detection by near-Earth spacecraft. The results appear to establish a physical link between coronal magnetic topology, formation of type III radio bursts, and production of SIEEs.
\end{abstract}


\section{Introduction}

Solar impulsive energetic electrons (SIEEs) are one population of solar energetic particles (SEPs) produced by the rapid release of magnetic energy on the Sun. In situ observations \citep{lin85,rea99} show that SIEEs are in close association with type III radio bursts. In some cases, the SIEEs are accompanied by enrichment of low-energy $^{3}$He (e.g. $^{3}$He/$^{4}$He $\geq$ 0.01) and enhancement of heavy ions (e.g. Fe). Traditionally, it was believed that solar flares rather than coronal mass ejections (CMEs) are the acceleration sources of impulsive SEPs, even though some unusual cases appear to be associated with narrow or jet-like CMEs \citep{wan06,pic06}. Recently, \citet{wan12} carried out a statistical survey of electron events over one solar cycle and found that only $\sim$35$\%$ of them are related to a reported flare, as compared to that of $\sim$60$\%$ that are related to a CME and to that of $\sim$50$\%$ for which these CMEs are narrow.

The narrow or jet-like CMEs can be explained by the interchange reconnection \citep{shi94,shi00} for which expanding loops (closed fields) reconnect with large scale open magnetic fields. Plasma jets are produced upward from the reconnection region along the open field lines and appear to be narrow CMEs extending to a few solar radii ($R_{s}$) in white light (WL) observations. Meanwhile, charged particles are accelerated and escape to the interplanetary space following the open field lines. Therefore, coronal magnetic topology indeed plays an important role in triggering solar eruptions and guiding energetic particles. Several authors \citep{wany06,pic06,rus08,mas09,li10,li11} have presented magnetic flux tubes rooted in solar active regions (ARs) that made it possible for energetic particles to easily escape from acceleration sites and quickly reach the Earth .

With this insight, it is necessary to examine the possibilities of solar sources or ARs in producing SEPs under different conditions of coronal magnetic fields. The impulsive SEP event that occurred on 2002 October 20 gives us an excellent opportunity to carry out this cross disciplinary study. This event had been partially discussed by several other authors \citep{wany06,pic06,wan11}. Unlike previous studies, we derive the solar particle release (SPR) times of SIEEs by considering the instrumental effect and find double injections of electrons at the Sun. We further investigate two candidate ARs that are potential sources of SIEEs and reveal a physical link between coronal magnetic topology, formation of type III radio bursts, and production of SIEEs.

\section{In situ particle measurements}

\begin{figure*}
     \centering
     \vspace{0.0\textwidth}    % Shift back to the panel bottom
     \centerline{
               \hspace*{0.00\textwidth}
               \includegraphics[width=0.33\textwidth]{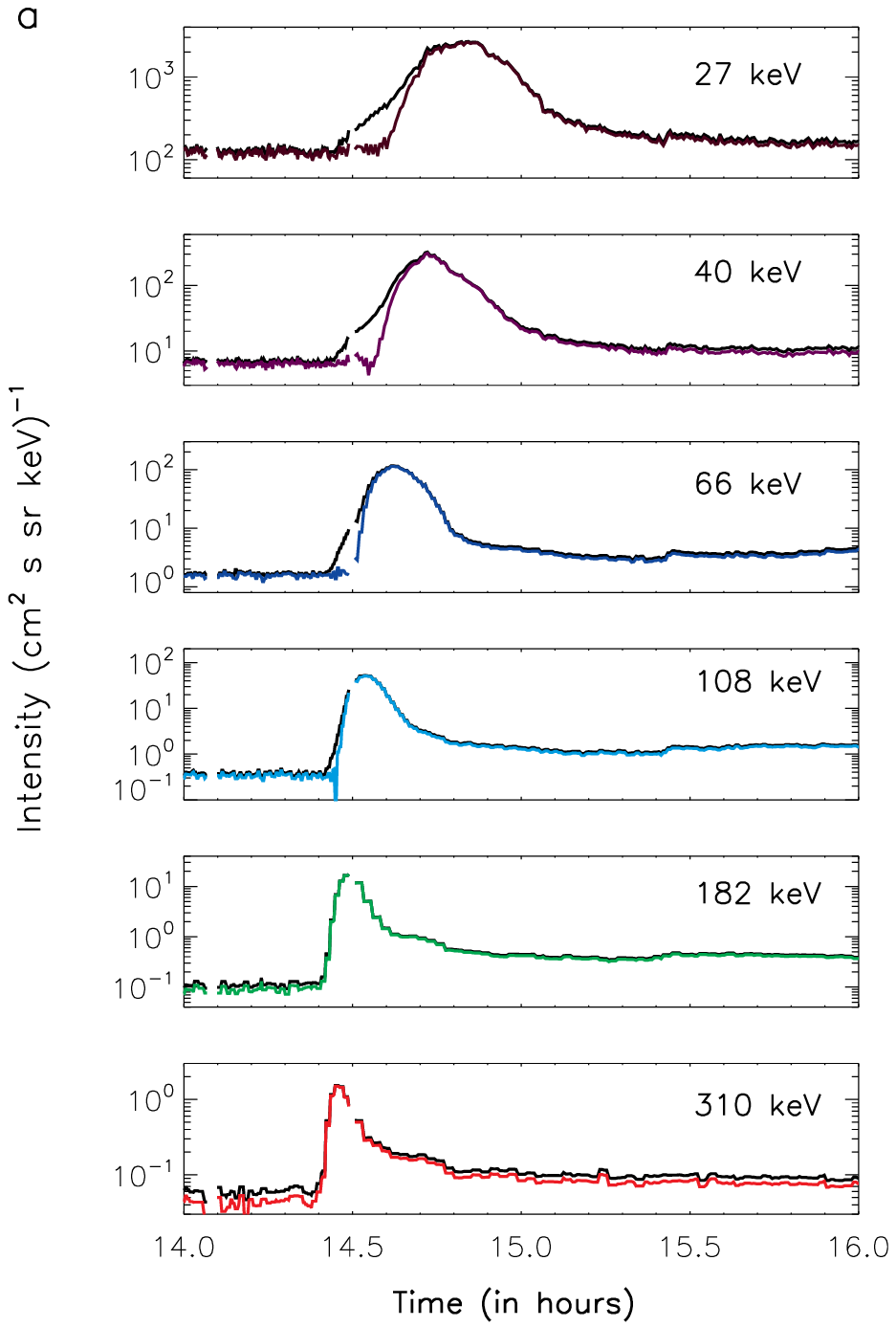}
               \includegraphics[width=0.32\textwidth]{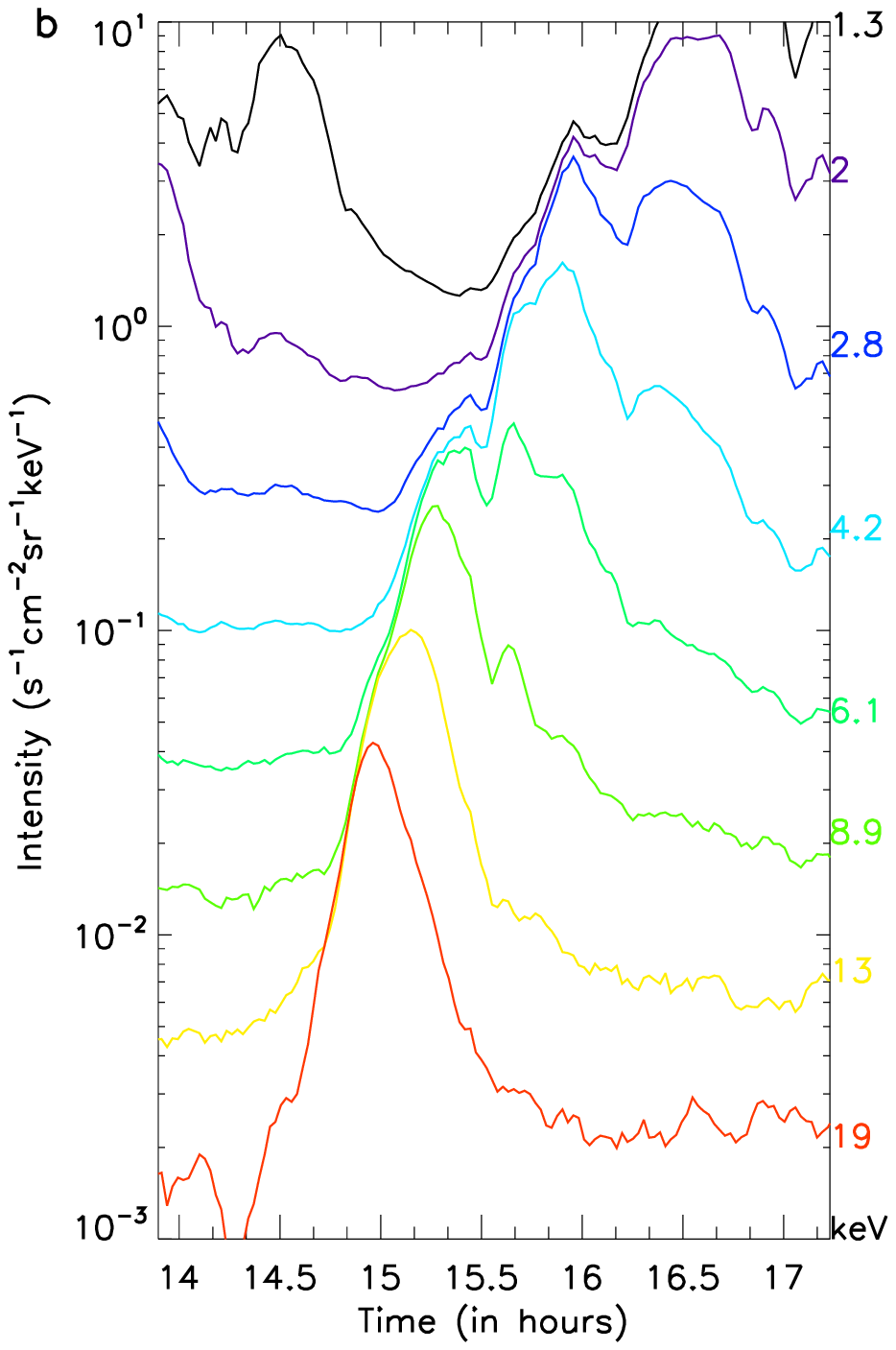}
               \includegraphics[width=0.32\textwidth]{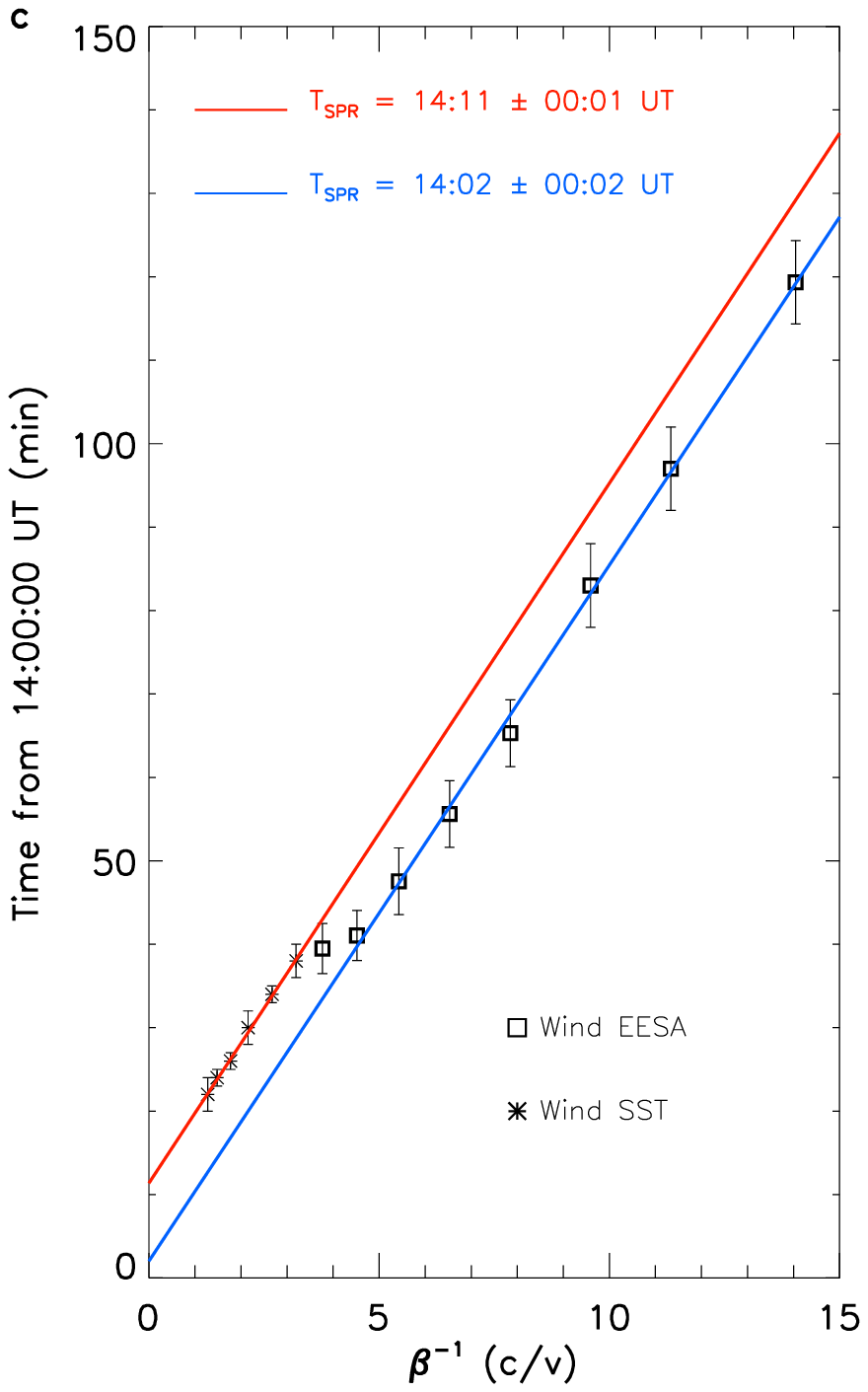}
               }
     \vspace{0.0\textwidth}   % Shift close to the panel top
\caption{Panel (a): Electron intensity profiles detected by $\sl WIND$/3DP/SSTs (27 -- 310 keV) on 2002 October 20. The black lines indicate the original fluxes, and the colored lines show fluxes after correction for scatter-out electrons. Panel (b): Electron intensity profiles detected by $\sl Wind$/3DP/EESAs (1.3 -- 19 keV). Panel (c): Onset times of electron fluxes as a function of inverse velocities. The red line indicates a linear fit to electrons above $\sim$25 keV, and the blue one shows a fit below 13 keV.}\label{fig1}
\end{figure*}

We use particle data from two near-Earth spacecraft, $\sl Wind$ and $\sl Advanced\ Composition\ Explorer$ ($\sl ACE$). The electron data have been obtained with the three-dimensional plasma and energetic particles instrument \citep[3DP;][]{lin95} onboard the $\sl Wind$. The electrostatic analyzers (EESAs) detect electrons from $\sim$0.5 to 28 keV, and the solid-state telescopes (SSTs) detect from 27 to 517 keV. The $\sl ACE$ Ultra Low Energy Isotope Spectrometer \citep[ULEIS;][]{mas98} provides elemental and isotropic measurements of He -- Ni from $\sim$0.02 to 10 MeV nucleon$^{-1}$.

For $\sl Wind$/3DP/SSTs, a laboratory calibration shows that a proportion of incident electrons will scatter back out of each detector before fully depositing their original energy and produce secondary particles that contaminate lower energy channels. This leads to erroneous results when studying the electron release from the Sun, if the contamination is not considered \citep{wan09}. An empirical method is applied here to correct the instrumental contamination \citep[for more details, refer to][]{sun12,li13}.

Figure 1 (panel a) presents the electron intensity profiles observed by $WIND$/3DP/SSTs in the energy range of 27 to 310 keV before (black lines) and after (color lines) correction on 2002 October 20. The velocity dispersion (onset times are later for lower energies) of in situ electrons is clearly shown, and the intensity maxima is found to be within 15 minutes from the onset time of in situ electrons. These indicate both solar impulsive injection and interplanetary scatter-free propagation of SIEEs. Figure 1 (panel b) also presents the intensity profiles of low-energy (1.3 -- 19 keV) electrons observed by $WIND$/3DP/EESAs. An earlier event occurred a few hours before. It is counted in low-energy channels.

A linear fit to the velocity dispersion has been commonly applied to study the solar particle release (SPR) time of SIEEs \citep{lin81,rea85,kru99}, assuming scatter-free propagation in interplanetary medium (IPM) for the first arriving electrons. The SPR time of electrons with energy $E_{n}$ can be expressed as $T_{SPR}(E_{n})=T_{onset}(E_{n})-L/v(E_{n})$, where $T_{onset}(E_{n})$ is the onset time of the intensity increase at 1 AU for the energy channel number $n$, $L$ is the interplanetary magnetic field (IMF) path length from solar release site to the near-Earth space, and $v(E_{n})$ is the electron velocity of the corresponding energy channel.

\begin{figure}
   \centering
   \vspace{-0.05\textwidth}
   \includegraphics[width=17pc]{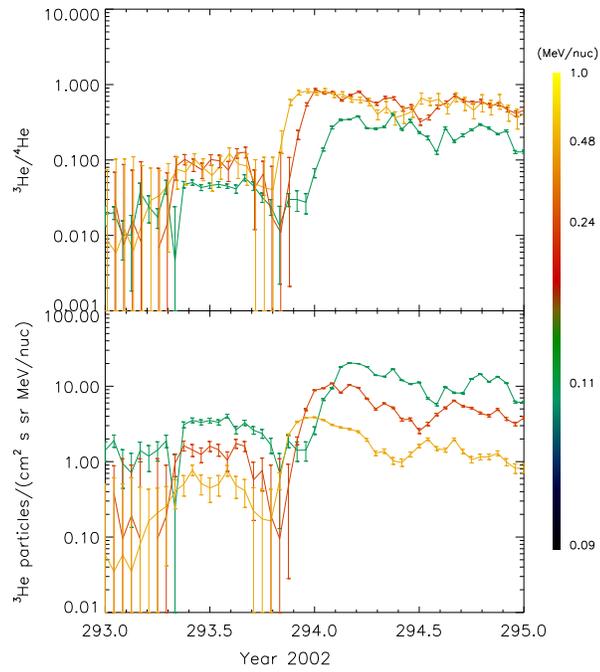}
   \vspace{-0.03\textwidth}
      \caption{The intensity profiles of $^{3}$He (bottom panel) and ratios of $^{3}$He/$^{4}$He (top panel) observed by $\sl ACE$/ULEIS. The energy ranges for the blue, red, and yellow lines are 0.06 -- 0.16, 0.16 -- 0.32, and 0.32 -- 0.64 MeV nucleon$^{-1}$, respectively.}
         \label{fig2}
\end{figure}

We can then plot $T_{onset}(E_{n})$ versus $c/v(E_{n})$, as shown in Figure 1 (panel c). It leads to a linear fit, where the y-axis intercept corresponds to the SPR time and the slope to the IMF path length. Note that the linear fit to the low-energy ($\lesssim$13 keV) channels leads to the electron SPR time being 14:02 $\pm$ 00:02 UT, while whose for high-energy ($\sim$25 to $\sim$300 keV) are 14:11 $\pm$ 00:01 UT. This suggests that there exist double injections for two species of electrons. More evidence is that both fits show similar slopes, indicating the two species of electrons travel along the same IMF path length of 1.03 $\pm$ 0.02 AU. This is slightly smaller than the nominal IMF line length of $\sim$1.06 AU, derived from the solar wind speed of $\sim$650 km $\rm s^{-1}$ during this event.

This event is in association with a variety of ion emissions, especially the enrichment of $^{3}$He. Figure 2 shows the intensity profiles of $^{3}$He (bottom panel) and ratios of $^{3}$He/$^{4}$He (top panel), as observed by the $\sl ACE$/ULEIS. It shows again clear velocity dispersion, and the maximum $^{3}$He/$^{4}$He ratio is $\sim$1.0 at 0.32 -- 0.64 MeV nucleon$^{-1}$. Both are the typical features of an impulsive SEP event. Note that the extrapolated low-energy ion injection at the Sun starts at 14:47 $\pm$ 00:33 UT \citep{pic06}, and the injection for protons is at 14:12 $\pm$ 00:05 UT \citep{kle05}, which is more compatible with the injection of the high-energy ($\sim$25 to $\sim$300 keV) electrons rather than those of low-energy ($\lesssim$13 keV).

\section{Solar observations and coronal magnetic topologies}

\begin{figure}
   \centering
   \vspace{0.0\textwidth}
   \includegraphics[width=22pc]{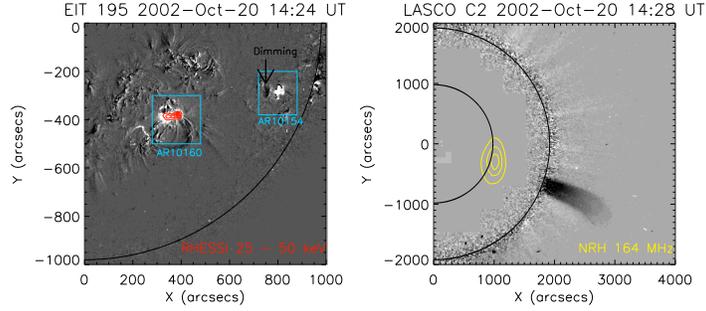}
   \vspace{-0.03\textwidth}
      \caption{Left panel: $\sl SOHO$/EIT 195 $\rm {\AA}$ image with a previous one subtracted. Red contour lines indicate the $\sl RHESSI$ HXR source in 25--50 keV at 14:22:20 UT. Blue boxes mark AR10154 and AR10160, respectively. Black arrow marks the dimming region. Right panel: Jet-like CME observed by $\sl SOHO$/LASCO WL image with a previous one subtracted. Yellow contour lines indicate the NRH radio source in 164 MHz at 14:12:09 UT.}
         \label{fig3}
\end{figure}

To clarify the acceleration sources of the SIEEs, especially for the two species of electrons, we combine solar observations by two instruments onboard the $\sl Solar\ and\ Heliospheric\ Observatory$ ($\sl SOHO$). Figure 3 shows the 195 $\rm {\AA}$ difference image with a previous one subtracted from the Extreme-Ultraviolet Imaging Telescope \citep[EIT;][]{del95} and the WL difference image from the Large Angle and Spectrometric Coronagraph \citep[LASCO;][]{bru95}. It is clear that two ARs relating to solar flares and a jet-like CME are identified to be the candidate acceleration sources of SIEEs.

AR10154 was positioned at S14W62 on the Sun and produced a C6.6 flare. The difference image in 195 $\rm {\AA}$ clearly shows a dimming region after the solar eruption (Figure 3, left panel). Note that the dimming is temporally and spatially in accordance with a jet-like CME that extends to several $R_{s}$ as shown by the WL difference image (right panel). Moreover, the jet-like CME is aligned with the 164 MHz radio source that is observed by the Nan\c{c}ay Radioheliogragh (NRH). With a larger active area, AR10160, was positioned at S06W25 and produced an C1.7 and an M1.8 flare. The M1.8 flare is associated with a contoured hard X-ray (HXR) source (left panel), which is reconstructed from the Reuven Ramaty High Energy Solar Spectroscopic Imager \citep[$\sl RHESSI$;][]{lin02}. It was, however, not associated with a reported CME, indicating that the flare was confined and failed to produce mass eruptions.

\begin{figure}
   \centering
   \vspace{0.0\textwidth}
   \includegraphics[width=12pc]{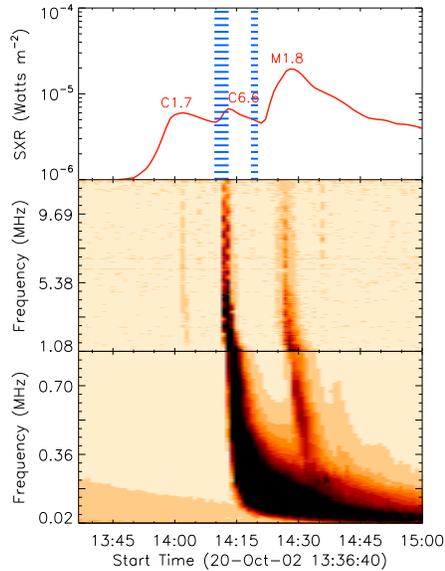}
   \vspace{0.02\textwidth}
      \caption{$\sl GOES$ SXR flux in 1 -- 8 $\rm {\AA}$ and the $\sl WIND$/WAVES radio dynamic spectra in the frequency range of 20 kHz to 14 MHz. Blue dashed lines mark the SPR times of low-energy ($\lesssim$13 keV) and high-energy ($\sim$25 to $\sim$300 keV) electrons, respectively. Line thickness indicates the error estimate.}
         \label{fig4}
\end{figure}

Figure 4 shows the three successive solar flares recorded by the $GOES$ soft X-ray (SXR) in 1 -- 8 $\rm {\AA}$ and the accompanied three groups of type III radio bursts observed by the WIND/WAVES radio spectrograms in the frequency range from 20 kHz to 14 MHz. Two vertical dashed lines mark the SPR times (plus 8.3 min with respect to electromagnetic emission) of the electrons with energies $\lesssim$13 keV and those ranging from $\sim$25 to $\sim$300 keV. It is found that the low-energy electrons injection coincides with the C6.6 flare in AR10154 and the prominent type III radio bursts, rather than the M1.8 flare from the relatively strong AR10160 that produced faint type III radio bursts. The high-energy electrons are, however, released $\sim$9 min later, when the jet-like CME travels to $\sim$2.6 $R_{s}$, according to the LASCO C2 observations (see Figure 3, right panel).

The high flux of in situ electrons and the prominent type III radio bursts may indicate that a great portion of open field lines are rooted in AR10154, while closed field lines dominate above AR10160. Therefore, it is necessary to examine the magnetic field configurations above the two ARs. We trace the coronal magnetic field lines, rooted in the two ARs and extended to 2.5 $R_{s}$, by applying the potential field source surface (PFSS) model \citep{sch03}. We then estimate the connection longitudes of the spiral IMF lines that connect the Sun with the Earth. This is based on the equation $\phi \simeq \omega r/u$, where $\omega$ is the angular speed of solar rotation, $r$ is the radial Sun-Earth distance, and $u$ is the solar wind speed. Figure 5 shows the extrapolated coronal magnetic topologies above AR10154 and AR10160 with open field lines marked in red and closed ones in brown, respectively. The connection longitudes with uncertainties of $\pm$10$^{\circ}$ are indicated by black lines.

It is evident that open field lines dominate above AR10154. Therefore, low-energy ($\lesssim$13 keV) electrons accelerated by the C6.6 flare can easily escape from the low corona to interplanetary space and generate the prominent type III radio bursts. This agrees with the Langmuir waves that scatter to produce the type III radio emission that were observed simultaneously with the arrival of $\sim$2 -- 10 keV electrons at 1 AU \citep{erg98}. The open field lines are well connected to the IMF lines that connect the Sun with the Earth, facilitating electrons to quickly reach the spacecraft. Note that these lines are also aligned with the jet-like CME (see Figure 3, right panel), which may be responsible for the production of high-energy ($\sim$25 to $\sim$300 keV) electrons. Whereas only a small portion of open field lines are rooted in AR10160. This explains why the stronger M1.8 flare produces relatively faint type III radio bursts. These lines are, however, not well connected to Earth, making it impossible for SEPs detection by near-Earth spacecraft.

\begin{figure}
\centering
\vspace{0.0\textwidth}
\noindent\includegraphics[width=20pc]{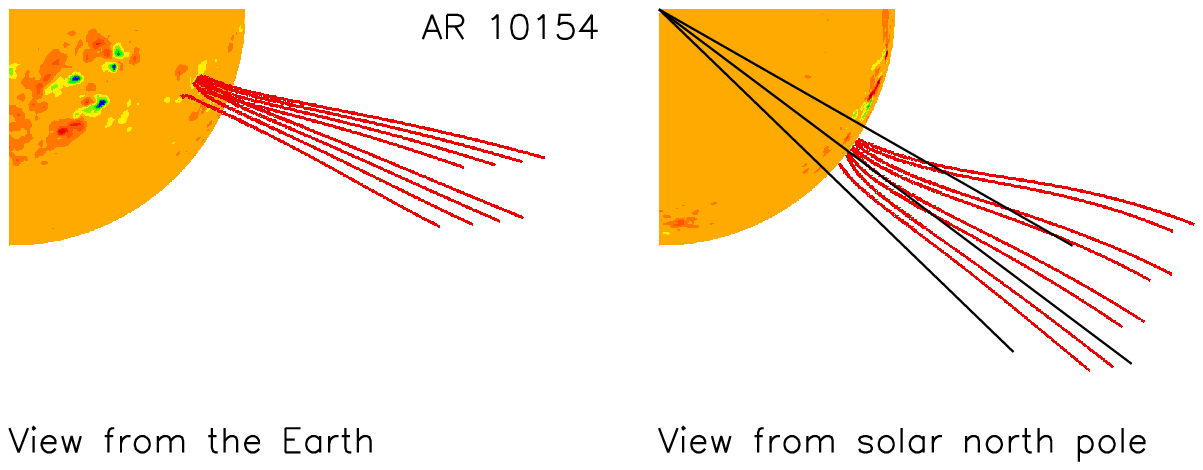}\\
\vspace{-0.02\textwidth}
\noindent\includegraphics[width=20pc]{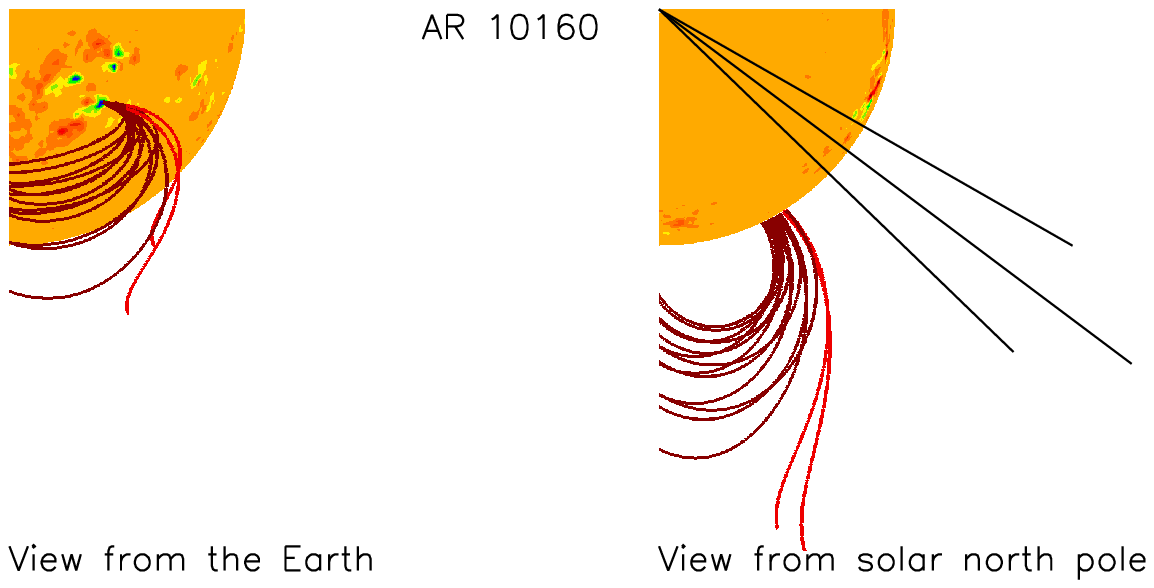}\\
\vspace{0.0\textwidth}
\caption{PFSS modeled coronal magnetic configurations above AR10154 and AR10160, respectively. The red lines mark the open field lines and brown lines mark the closed ones. Black lines mark the connection longitudes where the well connected IMF lines intersect the solar source surface. \label{fig5}}
\end{figure}

\section{Summary and discussion}

We re-investigate the 2002 October 20 SEP event by combining in situ particle measurements, remote sensing solar observations, and the magnetic field extrapolations. Data analysis leads to the following results and our main conclusions: (1) The SPR time of low-energy ($\lesssim$13 keV) electrons is derived as 14:02 $\pm$ 00:02 UT, which coincides with a C6.6 flare in AR10154. These electrons are probably the source of the prominent type III radio bursts. (2) Electrons of $\sim$25 to $\sim$300 keV are released $\sim$9 min later, when the jet-like CME travels to $\sim$2.6 $R_{s}$. (3) The open field lines, rooted in AR10154, are aligned with the jet-like CME and well connected with the IMF lines that connect the Sun with the Earth, providing escape channels for energetic electrons. (4) Only a small portion of magnetic fields are opened above AR10160, which is responsible for the faint type III radio bursts. These lines are, however, not well connected. Therefore, no SIEEs from AR10160 are detected by near-Earth spacecraft. (5) A physical link is evident between the coronal magnetic topology, the formation of type III radio bursts, and the production of SIEEs.

The time delay of high-energy electron injection at the Sun has been reported by several authors \citep{kru99,hag02,wan06}. For instance, \citet{wan06} found two distinct injections in three impulsive electron events: Electrons of $\sim$0.4 $-$ 9 keV are released earlier than type III radio emission, while the injection of $\sim$13 $-$ 300 keV electrons starts when the associated CME reaches $\sim$1 to 6 $R_{s}$. The delayed injection of high-energy electrons can be due either to the acceleration by CME-driven shocks or to the propagation effects in the IPM. \citet{wan11} further studied the pitch-angle distributions (PADs) of in situ electrons and found that high-energy electrons experience obvious pitch-angle scattering. This seems to support the propagation effects in the IPM. However the high-energy electron injection is derived to be compatible with the low-energy ions, especially for protons in the present study. We, therefore, cannot ignore the role of shock waves associated with CMEs in producing high-energy electrons. The low-energy electrons from flare sites can serve as seed particles for further acceleration by CME-driven shocks.

\begin{acknowledgements}
We thank the $\sl WIND$, $\sl ACE$, $\sl GOES$, $\sl SOHO$, and $\sl RHESSI$ spacecraft, as well as the Nan\c{c}ay Radioheliogragh for providing observational data. This work is supported by 985 project of Nanjing University and Advanced Discipline Construction Project of Jiangsu Province. C. Li wound like to acknowledge the Natural Science Foundation (BK2012299) of Jiangsu province, the Ph.D. programs foundation of Ministry of Education of China (20120091120034), and the Technology Foundation for Selected Overseas Chinese Scholar, Ministry of Personnel of China. X. Wang is supported by National Natural Science Foundation of China (NSFC 41074123).
\end{acknowledgements}

\end{document}